\documentclass[journal,10pt]{IEEEtran}
\usepackage{graphicx}
\usepackage{epstopdf}
\usepackage{float}
\usepackage{amsfonts}
\usepackage{amssymb}
\usepackage{amsmath}
\usepackage{accents}
\usepackage{multirow}
\usepackage{array}
\usepackage{pgfplots}
\usepackage{booktabs}
\usepackage{multirow}
\usepackage{epstopdf,epsfig}

\definecolor{mycolor1}{rgb}{0.00000,0.20000,0.60000}%
\definecolor{mycolor2}{rgb}{0.00000,0.40000,0.80000}%
\definecolor{mycolor3}{rgb}{0.40000,0.60000,1.00000}%
\definecolor{mycolor4}{rgb}{0.00000,0.60000,1.00000}%
\definecolor{mycolor5}{rgb}{1.00000,0.40000,0.60000}%
\definecolor{mycolor6}{rgb}{1.00000,0.20000,0.40000}%
\definecolor{mycolor7}{rgb}{1.00000,0.00000,0.20000}%
\definecolor{mycolor8}{rgb}{0.60000,0.20000,0.00000}%
\ifCLASSINFOpdf
\else
\fi
\begin{document}

\title{Study of Linear Precoding and Stream Combining for Rate Splitting in MU-MIMO Systems \vspace{-0.05em}}

%
\author{André Flores, Rodrigo C. de Lamare and Bruno Clerckx \\
}

\maketitle

\begin{abstract}
This paper develops stream combining techniques for rate-splitting (RS) multiple-antenna systems with multiple users to enhance the common rate. We propose linear combining techniques based on the Min-Max, the maximum ratio and the minimum mean-square error criteria along with Regularized Block Diagonalization (RBD) precoders for RS-based multiuser multiple-antenna systems. An analysis of the sum rate performance is carried out, leading to closed-form expressions. Simulations show that the proposed combining schemes offer a significant sum rate performance gain over conventional linear precoding schemes. 
\end{abstract}

\begin{IEEEkeywords}
Multiuser MIMO, ergodic sum rate, rate-splitting, regularized block diagonalization.
\end{IEEEkeywords}

%
\IEEEpeerreviewmaketitle

\section{Introduction}
Multiuser multiple-input multiple-output (MIMO) systems can provide high data rates through spatial multiplexing to distributed users. However, multiuser interference (MUI) can heavily degrade the overall performance of a MIMO system [1]. Therefore, several precoding techniques aiming to mitigate the MUI have been reported in the literature \cite{Joham2005,Spencer2004}. The main drawback of these methods is that they rely on very accurate channel state information at the transmitter (CSIT), which remains challenging to acquire in practice\cite{JoudehClerckx2016}.

In the past years, rate-splitting (RS) has been established as a promising strategy to improve the performance of Multiuser MIMO, even under imperfect CSIT  \cite{Clerckx2016}. RS splits original messages into common and private parts, encodes the common and privates parts into streams, precode and then transmit them in a superimposed manner. At the receivers, all users use successive interference cancellation (SIC) to decode and cancel the common stream, before each user can decode its private stream. The key advantage of RS is the flexibility introduced by the split of the messages and the creation of the common stream, whose content and power can be adapted with the purpose of adjusting how much interference is canceled by the receivers. This enables to flexibly manage multiuser interference between the two extremes of fully decode interference and fully treat it as noise \cite{Maoinpress}. 

RS has been used in \cite{JoudehClerckx2016} and \cite{Hao2015} with linear precoders and in \cite{Flores2018} with non-linear precoders considering perfect and imperfect CSIT. RS for robust transmission has been studied in \cite{Joudeh2016a}. In \cite{LuLiTianEtAl2018}, RS has been implemented to reduce the effects of the imperfect CSIT caused by finite feedback. However, previous works focus on multiple-input single-output (MISO) systems along with either optimized or closed-form zero-forcing (ZF) and minimum mean-squared error (MMSE) channel inversion-type precoders. MIMO systems, with multiple receive antennas, have been considered in \cite{HaoB.Clerckx2017} from a Degrees-of-Freedom (DoF) perspective. In \cite{Flores2019}, RS has been employed in a MIMO scenario using the BD precoder and two different common stream combining techniques. The results show that the common combiner has the potential to significantly increase the sum rate of MIMO systems. 


In this work, we present stream combining techniques along with regularized block diagonalization (RBD) precoder for RS in multiuser MIMO systems \cite{streamcomb}. We consider a different receiver structure than the one employed in \cite{Flores2019} in order to simplify the combiners and reduce the computational complexity. We also propose the MMSE common stream combiner to further enhance the rate of the common stream and compared its performance with the Min-Max and Maximum Ratio stream combiners. We derive closed form expressions to describe the sum rate performance of the proposed schemes. Furthermore, analytical expressions for the sum rate of the proposed combiners with the RBD precoder are derived. Simulations assess the performance of the proposed approaches against existing techniques under both perfect and imperfect CSIT.

The rest of this paper is organized as follows. Section II describes
the system model and reviews the RBD
precoding technique. Section III presents the proposed combining strategies and the structure of the receiver. In Section
IV, the analysis of the sum rate performance is carried out. The simulation results are displayed in Section V. 
Finally, Section VI concludes this work.

Matrices and vectors are represented by upper and lowercase boldface letters respectively. The conjugate transpose of a matrix is denoted by $\left(\cdot\right)^{H}$, whereas $\left(\cdot\right)^{\text{T}}$ denotes the transpose of a matrix. The operators $\lVert \cdot \rVert$, $\odot$, and $\mathbb{E}\left[\cdot\right]$ stand for the Euclidean norm, the Hadamard product and the expectation operator. The trace of a matrix and the cardinality of a set are given by $\text{tr}\left(\cdot\right)$, and $\text{card}\left(\cdot\right)$. $\text{diag}\left(\mathbf{c}\right)$ creates a diagonal matrix with the entries of $\mathbf{c}$ in the main diagonal. 
\section{System Model and Linear Precoding} 

\begin{figure}[htb]
\begin{centering}
\includegraphics[scale=0.4]{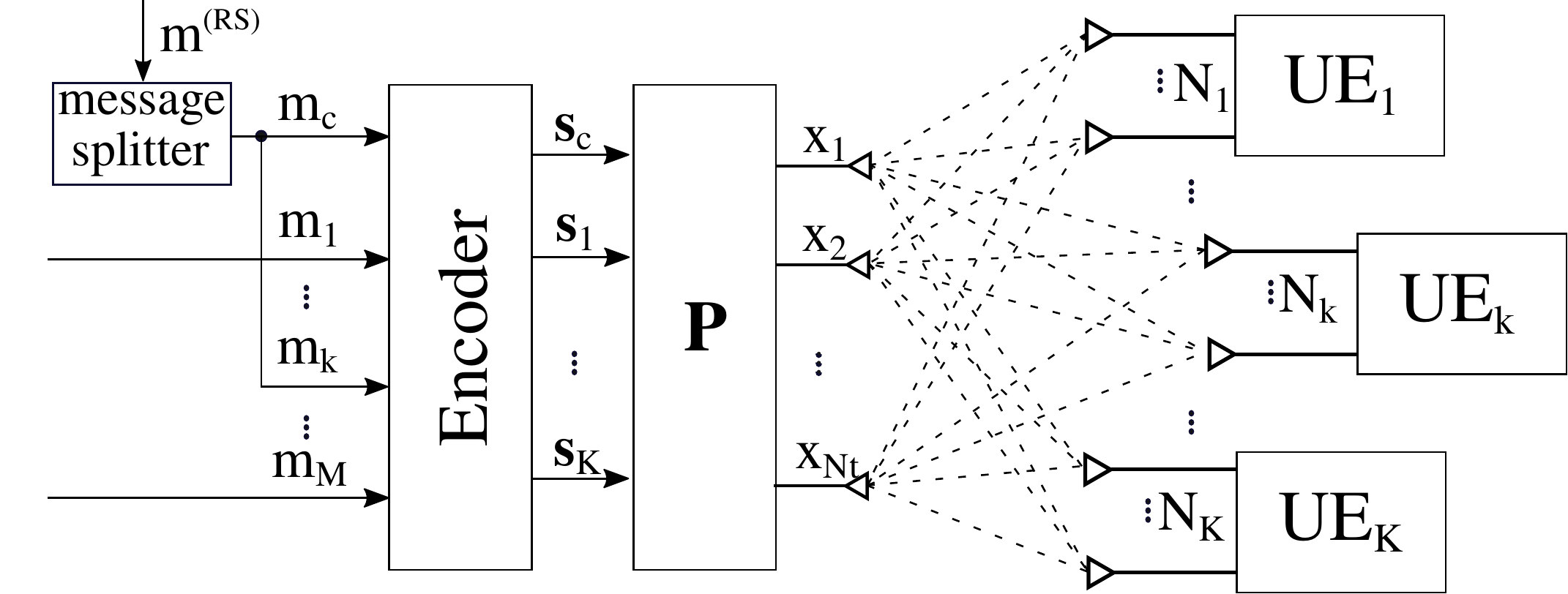}
\par\end{centering} \vspace{-0.75em}
\caption{System model.}\label{System Model}
\end{figure}

Let us consider the downlink of a MIMO system with $K$ users, as depicted in Fig. \ref{System Model}. The $k$th User Equipment (UE) is equipped with $N_k$ antennas i.e., the total number of receive antennas is $N_r=\sum_{k=1}^{K}N_k$. The transmitter has $N_t$ antennas, where $ N_t \geq K \geq 2$. We consider RS in a system where the BS wants to deliver $M$ messages to the users, and, for simplicity, splits only one message into a common part and a private part, e.g. message $\text{m}^{\left(\text{RS}\right)}$ as in Fig. 1. The BS then encodes $1$ common part and $M$ private parts (the private part from $\text{m}^{\left(\text{RS}\right)}$, namely $\text{m}_k$, and the remaining $M-1$ messages that have not been split), similarly to \cite{JoudehClerckx2016,Clerckx2016,Maoinpress,Hao2015,Flores2018,Joudeh2016a,LuLiTianEtAl2018}. The set $\mathcal{M}_k$ contains the data streams of the $k$th user. The number of data streams transmitted is equal to $M+1=\sum_{k=1}^{K} M_k+1$ with $M_k=\text{card}\left(\mathcal{M}_k\right)$ and $n_k=\sum_{j=1}^{k-1} M_j$.  

The data stream $\text{m}^{\left(\text{RS}\right)}$ is split and then modulated, resulting in a vector of symbols $\mathbf{s}^{\left(\text{RS}\right)}=\left[s_c,\mathbf{s}_1^{\text{T}},\mathbf{s}_2^{\text{T}},\dots,\mathbf{s}_K^{\text{T}}\right]^{\text{T}} \in \mathbb{C}^{M+1}$. The common symbol is denoted by $s_c$, whereas the vector $\mathbf{s}_k$ contains the $M_k$ private streams of the $k$th user. We assume that the symbols have zero mean and covariance matrix equal to the identity matrix.

At the transmitter, a linear precoder $\mathbf{P}^{\left(\text{RS}\right)}=\left[\mathbf{p}_c,\mathbf{P}\right]\in\mathbb{C}^{N_T\times \left(M+1\right)}$ maps the symbols to the transmit antennas. In particular, $\mathbf{p}_c \in \mathbb{C}^{N_t}$ performs the mapping of the common symbol\footnote{Receivers with multiple antennas allow the transmission of a vector of common symbols/streams, which could further enhance the performance \cite{HaoB.Clerckx2017}. This is left for further studies.} to the transmit antennas. The private precoder is given by $\mathbf{P}=\left[\mathbf{P}_1,\mathbf{P}_2,\dots,\mathbf{P}_K\right]$, where $\mathbf{P}_k\in \mathbb{C}^{N_t\times M_k}$ denotes the private precoder of the $k$th user and the vector $\mathbf{p}_k$ denotes the $k$th column of $\mathbf{P}$.

Let us consider a general diagonal power loading matrix $\mathbf{A}^{\left(\text{RS}\right)}=\text{diag}\left(\mathbf{a}^{\left(\text{RS}\right)}\right) \in \mathbb{R}^{\left(M+1\right)\times \left(M+1\right)}$. The vector $\mathbf{a}^{\left(\text{RS}\right)}=[a_c, \mathbf{a}_1^{\text{T}}, \mathbf{a}_2^{\text{T}},\cdots,\mathbf{a}_k^{\text{T}}]^{\text{T}}$ assigns the power to the common and private streams. Specifically, the vector $\mathbf{a}_k \in \mathbb{R}^{M_k}$ allocates the power to the $M_k$ private symbols in $\mathcal{M}_k$ and the coefficient $a_c$ designates the power to the common message. Then, the transmitted signal is expressed by 
\begin{equation}
\mathbf{x}=a_c s_c\mathbf{p}_c +\sum_{k=1}^K\mathbf{P}_k\text{diag}\left(\mathbf{a}_k\right)\mathbf{s}_k.\label{transmit signal}
\end{equation}
The model satisfies the transmit power constraint
$\mathbb{E}\left[\lVert\mathbf{x}\rVert^2\right]\leq E_{tr}$, where
$E_{tr}$ denotes the total transmit power. 
The transmit vector $\mathbf{x}$ passes through the channel $\mathbf{H}^{\text{T}}=\mathbf{\hat{H}}^{\text{T}}+\mathbf{\tilde{H}}^{\text{T}} \in \mathbb{C}^{N_r\times N_t}$, where $\mathbf{\hat{H}}^{\text{T}}$ designates the channel estimate and the matrix $\mathbf{\tilde{H}}^{\text{T}}$ models the CSIT quality by adding the estimation error. The matrix $\mathbf{H}_k^{\text{T}}\in\mathbb{C}^{N_k\times N_t}$ represents the channel of the $k$th user. It follows that $\mathbf{H}= [{\mathbf H}_1, \ldots, {\mathbf H}_k, \ldots, {\mathbf H}_K]$. 
 For simplicity, we consider a flat fading channel which remains fixed during a transmission block. 

The signal obtained at the $k$th user following \eqref{transmit signal} is  
\begin{align}
\mathbf{y}_k=&\overbrace{a_c s_c \mathbf{H}_k^{\text{T}}\mathbf{p}_c}^{\text{Common stream}} +\overbrace{\sum_{i\in\mathcal{M}_k}a_i s_i\mathbf{H}_k^{\text{T}}\mathbf{p}_i}^{\text{User-k private streams}}+\overbrace{\sum_{\substack{j=1\\j\notin\mathcal{M}_k}}^{M} a_js_j\mathbf{H}_k^{\text{T}}\mathbf{p}_j}^{\text{Multi-User Interference}}+\mathbf{n}_k,\label{Received signal RS user k}
\end{align}
where the noise vector $\mathbf{n}_k\in\mathbb{C}^{N_k\times 1}$ is assumed uncorrelated and follows a complex normal distribution i.e., $\mathbf{n}_k\sim
\mathcal{CN}\left(\mathbf{0},\sigma_n^2\mathbf{I}\right)$. 
The power of \eqref{Received signal RS user k} at the $i$th receive antenna is given by 
\begin{align}
\mathbb{E}\left[\lvert \text{y}_{k,i}\rvert^2\right]=&a_c^2  \lvert\mathbf{h}^{\left(k\right)^{\text{T}}}_{i}\mathbf{p}_c\rvert^2+\sum_{\substack{j=1}}^M a_j^2\lvert\mathbf{h}^{\left(k\right)^{\text{T}}}_{i}\mathbf{p}_j\rvert^2+\sigma_n^2.\label{Received signal RS antenna k}
\end{align}
Under perfect CSIT assumption, the estimation error goes to zero and equations \eqref{Received signal RS user k} and \eqref{Received signal RS antenna k} remain the same with $\mathbf{H}=\mathbf{\hat{H}}$. Note that the conventional non-RS MIMO system represents a particular case of the model established where no power is distributed to the common message, i.e., $a_c=0$ (and therefore no split of the message is conducted).


In what follows we consider the RBD precoding technique \cite{StankovicHaardt2008,SungLeeLee2009,lclattice,glcbd,wlbd} to define the private precoder.
RBD separates the precoder into two matrices, i.e., $\mathbf{P}_{k}^{\left(\text{RBD}\right)}=\mathbf{P}^a_k\mathbf{P}^b_k.$
The filter $\mathbf{P}^a_k$ partially removes MUI \cite{StankovicHaardt2008} and is computed through the following optimization problem:
\begin{equation}
\mathbf{P}^a_k=\min_{\mathbf{P}^a_k}\mathbb{E}\left[\lVert\mathbf{\bar{H}}^{\text{T}}_k\mathbf{P}^a_k\rVert^2+\frac{\lVert\mathbf{n}_k\rVert^2}{\delta^2}\right],\label{RBD optimization}
\end{equation}
where the matrix $\mathbf{\bar{H}}_k$ is formed by excluding the $k$th user, i.e. $\mathbf{\bar{H}}_k=\left[\mathbf{H}_1, \ldots, \mathbf{H}_{k-1}, ~\mathbf{H}_{k+1}, \ldots, \mathbf{H}_K\right]$ and the parameter $\delta$ is a scaling factor imposed in order to fulfil the transmit power constraint. By applying SVD we get $\mathbf{\bar{H}}_k^{\text{T}}=\mathbf{\bar{U}}_k\mathbf{\bar{\Psi}}_k\mathbf{\bar{V}}_k^H$. The solution to \eqref{RBD optimization} is given by
\begin{equation}
\mathbf{P}^a_k=\mathbf{\bar{V}}_k\left(\mathbf{\bar{\Psi}_k}^{\text{T}}\mathbf{\bar{\Psi}}_k+\frac{N_r\sigma_n^2}{E_{tr}}\mathbf{I}_{N_t}\right)^{-1/2}
\end{equation}

The second filter $\mathbf{P}^b_k$ allows parallel symbol detection. Consider the effective channel matrix defined as $\underaccent{\ddot}{\mathbf{H}}_{k}^{\text{T}}=\mathbf{H}_k^{\text{T}}\mathbf{P}^a_k$. A second SVD  is computed on the effective channel , i.e., $\underaccent{\ddot}{\mathbf{H}}_{k}^{\text{T}}=\underaccent{\ddot}{\mathbf{U}}_k \underaccent{\ddot}{\mathbf{\Psi}}_k\underaccent{\ddot}{\mathbf{V}}_k^H$, in order to find the second precoder and the receive filter of the $k$th user as given by 
\begin{align}
\mathbf{P}^b_k=&\underaccent{\ddot}{\mathbf{V}}_k, & \mathbf{G}^{\left(\text{RBD}\right)}_k=&\underaccent{\ddot}{\mathbf{U}}_k^H.\label{Precoder and receiver RBD}
\end{align}

\section{Proposed Stream Combining Techniques}

Let us consider a system employing an RS scheme with Gaussian signalling. The instantaneous common rate at the $k$th user is defined as 
\begin{equation}
R_{c,k}=\log_2\left(1+\gamma_{c,k}\right),\label{MRC common rate}
\end{equation}
where $\gamma_{c,k}$ is the Signal-to-Interference-plus-noise ratio (SINR) at the $k$th user when decoding the common message.

In order to evaluate the performance we consider the Ergodic Sum Rate (ESR) over a long sequence of fading channel states to ensure that the rates are achievable, as detailed in \cite{JoudehClerckx2016}. The total ESR of the system is given by
\begin{equation}
S_r=\min_{k\in \left[1,K\right]} \mathbb{E}\left[\bar{R}_{c,k}\right]+\mathbb{E}\left[\bar{R}_p\right]. \label{total ergodic rate}
\end{equation} 
The first term of \eqref{total ergodic rate} represents the ergodic common rate, where $\bar{R}_{c,k}=\mathbb{E}\left[R_{c,k}\lvert\hat{\mathbf{H}}\right]$. The min operator is used since all users should decode the common message. The second term denotes the ergodic sum-private rate with $\bar{R}_p=\mathbb{E}\left[R_p\lvert\hat{\mathbf{H}}\right]$. The sum-private rate $R_p$ embodies all private rates, i.e., $R_p=\sum_{k=1}^K R_k$, where $R_k$ denotes the instantaneous  private rate of the $k$th user. 
                                 
\begin{figure}[htb]
\vspace{-6mm}
\begin{centering}
\includegraphics[scale=0.4]{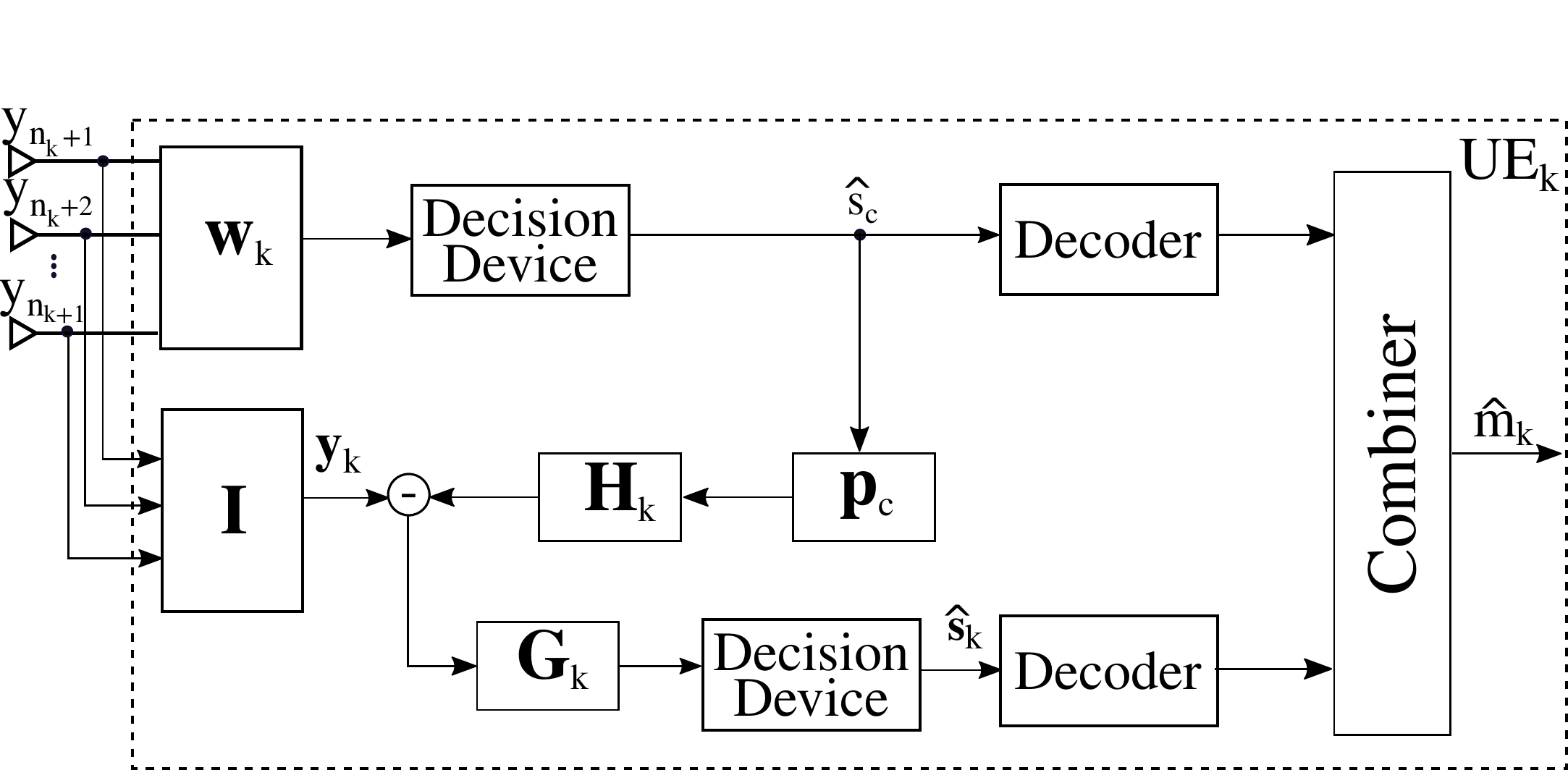}
\par\end{centering} \vspace{-0.75em}
\caption{Receiver structure.}\label{Receiver}
\end{figure}

Unlike receivers in RS MISO systems, the $k$th receiver in a MIMO system has access to $N_k$ copies of the common symbol. Let us consider  \eqref{Received signal RS user k}
 and define the combined signal $\tilde{\text{y}}_k=\mathbf{w}_k^H\mathbf{y}_k$, where the vector $\mathbf{w}_k=\left[w_1~~w_2~~\cdots~~w_{M_k}\right]^{\text{T}}$ represents the combining filter used to maximize the SNR. Let us define the vectors $\mathbf{r}_{k,c}=\mathbf{H}_k^{\text{T}}\mathbf{p}_c$ and $\mathbf{r}_{k,i}=\mathbf{H}_k^{\text{T}}\mathbf{p}_i$.  Then, the average power of $\tilde{\text{y}}_k$ is 
\vspace{-6mm}
\begin{equation}
\mathbb{E}\left[\lvert\tilde{\text{y}}_k\rvert^2\right]=a_c^2\lvert\mathbf{w}_k^{H} \mathbf{r}_{k,c}\rvert^2
+\sum_{j=1}^M a_j^2\lvert\mathbf{w}_k^{H}\mathbf{r}_{k,j}\rvert^2+\lVert\mathbf{w}_k\rVert^2\sigma_n^2.\label{MRC received signal power}
\end{equation}
\vspace{-1mm}
From \eqref{MRC received signal power} we get the common message SINR given by
\vspace{-1mm}
\begin{equation}
\gamma_{k,c}=\frac{a_c^2\lvert\mathbf{w}_k^{H}\mathbf{r}_{k,c}\rvert^2}{\sum\limits_{i\in\mathcal{M}_k}a_i^2\lvert\mathbf{w}_k^{H}\mathbf{r}_{k,i}\rvert^2+\sum\limits_{\substack{j=1\\j\notin \mathcal{M}_k}}^M a_j^2\lvert\mathbf{w}_k^{H}\mathbf{r}_{k,j}\rvert^2+\lVert\mathbf{w}_k\rVert^2\sigma_n^2}.\label{SINR general combiner}
\end{equation}
The structure of the $k$th receiver is shown in Fig. \ref{Receiver}, which is different from \cite{Flores2019}, where the combiner and the private receiver are implemented sequentially. In what follows, we propose combining strategies to set up $\mathbf{w}_k$ and enhance the common rate performance.

\subsection{Min-Max Criterion}
Let us consider \eqref{Received signal RS antenna k} from the model described in \eqref{Received signal RS user k}. The common rate obtained at the $i$th receive antenna of user $k$ can be computed by
\begin{equation}
R_{c,k,i}=\log_2\left(1+\frac{a_c^2\lvert\mathbf{h}_{i}^{\left(k\right)^{\text{T}}}\mathbf{p}_c\rvert^2}{\sum_{j=1}^M a_j^2\lvert\mathbf{h}_{i}^{\left(k\right)^{\text{T}}}\mathbf{p}_j\rvert^2+\sigma_n^2}\right).\label{individual Rc minmax criterion}
\end{equation}
The Min-Max criterion selects at each receiver the antenna that leads to the highest common rate, i.e., $R_{c,k}^{\left(\text{max}\right)}=\max_{i}\left(R_{c,k,i}\right). \label{MinMax criterion Max}$ The $k$th entry of $\mathbf{w}_k$ is set to one if the $k$th antenna is selected and all the other entries are set to zero. 
Using $R_{c,k}^{\left(\text{max}\right)}$ with \eqref{total ergodic rate} we get the sum rate performance. 

\subsection{Maximum Ratio Combining}
Another possibility to enhance the common rate is to use Maximum Ratio Combining (MRC). The maximum value of the numerator is achieved when $\mathbf{w}_k^{\left(\text{MRC}\right)}=\frac{\mathbf{r}_{k,c}}{\lVert \mathbf{r}_{k,c}\rVert^2} $ i.e., when the vectors $\mathbf{w}_k$ and $\mathbf{r}_{k,c}$ are parallel. Using the property of the dot product and simplifying terms in \eqref{SINR general combiner}, the SINR can be expressed as follows: 
\begin{equation}
\gamma_{k,c}^{\left(\text{MRC}\right)}=\frac{a_c^2\lVert \mathbf{r}_{k,c}\rVert^2}{\sum\limits_{i\in\mathcal{M}_k}a_i^2\lVert \mathbf{r}_{k,i}\rVert^2\cos\beta_i+\sum\limits_{\substack{j=1\\j\notin \mathcal{M}_k}}^M a_j^2\lVert \mathbf{r}_{k,j}\rVert^2\cos\beta_j+\sigma_n^2}\label{SINR MRC}.
\end{equation}
where $\beta_j$ is the angle between $\bold{w}_k$ and $\mathbf{r}_{k,j}$. The sum rate performance can be found by using \eqref{SINR MRC} in \eqref{MRC common rate} and in \eqref{total ergodic rate}.



\subsection{Minimum Mean-Square Error Combining}

The proposed MMSE combiner (MMSEc) considers the optimization problem given by
\begin{equation}
\mathbf{w}_k^{\left(\text{MMSE}\right)}=\min_{\mathbf{w}_k} \mathbb{E}\left[\lVert s_c-\mathbf{w}_k^{H}\mathbf{y}_k\rVert^2\right].\label{OptMMSEc}
\end{equation} 
Evaluating the expected value on the right side of \eqref{OptMMSEc}, we have
\begin{align}
\mathbb{E}\left[\lVert s_c-\mathbf{w}_k^{H}\mathbf{y}_k\rVert^2\right]=&\mathbb{E}\left[\left(s_c-\mathbf{w}_k^{H}\mathbf{y}_k\right)\left(s_c-\mathbf{w}_k^{H}\mathbf{y}_k\right)\right]\nonumber\\
=&1-\mathbf{w}_k^H\mathbf{H}_k^{\text{T}}\mathbf{p}_c-\mathbf{p}_c^H\mathbf{H}_k^*\mathbf{w}_k+\nonumber\\
&\mathbf{w}_k^H\mathbf{R}_{\mathbf{y}_k\mathbf{y}_k}\mathbf{w}_k,
\end{align}
where $\mathbf{R}_{\mathbf{y}_k\mathbf{y}_k}=\mathbb{E}\left[\mathbf{y}_k\mathbf{y}_k^H\right]$. Taking the derivative with respect to $\mathbf{w}_k^H$ and equating the result to zero we obtain
\begin{equation}
\frac{\partial\mathbb{E}\left[\lVert s_c-\mathbf{w}_k^{H}\mathbf{y}_k\rVert^2\right]}{\partial \mathbf{w}_k^H}=-\mathbf{H}_k^{\text{T}}\mathbf{p}_c+\mathbf{R}_{\mathbf{y}_k\mathbf{y}_k}\mathbf{w}_k=0.\label{dervativeMMSEc}
\end{equation}
Solving \eqref{dervativeMMSEc} with respect to $\mathbf{w}_k$ we get the MMSEc expression, which is given by
\begin{equation}
\mathbf{w}_k^{\left(\text{MMSE}\right)}=\mathbf{R}_{\mathbf{y}_k\mathbf{y}_k}^{-1}\mathbf{H}_k\mathbf{p}_c.
\end{equation}
Let us consider the quantities:
\begin{equation}
\lVert \mathbf{w}_k \rVert^2\sigma_n^2=\text{tr}\left(\mathbf{R}_{\mathbf{y}_k\mathbf{y}_k}^{-2}\mathbf{H}_k^{\text{T}}\mathbf{p}_c\mathbf{p}_c^H\mathbf{H}_k^*\right)\sigma_n^2,\label{MMSE noise}
\end{equation}
\begin{equation}
\lvert\mathbf{w}_k^H\mathbf{r}_{k,i}\rvert^2=\mathbf{p}_i^H\mathbf{H}_k^*\mathbf{R}_{\mathbf{y}_k\mathbf{y}_k}^{-1}\mathbf{H}_k^{\text{T}}\mathbf{p}_c\mathbf{p}_c^H\mathbf{H}_k^*\mathbf{R}_{\mathbf{y}_k\mathbf{y}_k}^{-1}\mathbf{H}_k^{\text{T}}\mathbf{p}_i,\label{MMSE symbols}
\end{equation}
\begin{equation}
\lvert\mathbf{w}_k^H\mathbf{r}_{k,c}\rvert^2=\mathbf{p}_c^H\mathbf{H}_k^*\mathbf{R}_{\mathbf{y}_k\mathbf{y}_k}^{-1}\mathbf{H}_k^{\text{T}}\mathbf{p}_c\mathbf{p}_c^H\mathbf{H}_k^*\mathbf{R}_{\mathbf{y}_k\mathbf{y}_k}^{-1}\mathbf{H}_k^{\text{T}}\mathbf{p}_c,\label{MMSE common}
\end{equation}
Substituting \eqref{MMSE noise},\eqref{MMSE symbols}, and \eqref{MMSE common} into \eqref{SINR general combiner} we obtain the SINR of MMSEc, which can be used with \eqref{MRC common rate} to get the common rate.
\vspace{-1mm}
\subsection{Private Rate}

The common symbol is removed from the received signal using a SIC technique \cite{spa,mfsic,jiomimo,mfdf,mbdf,bfidd}. A receive filter $\mathbf{G}_k\in\mathbb{C}^{M_k\times N_k}$ can be used to improve the detection of the private symbols. Let us consider the matrix $\mathbf{F}_k=\mathbf{G}_k\mathbf{H}_k^{\text{T}}$ in order to simplify the notation. Then, the achievable rate of the  $k$th user is described by 
\begin{equation}
R_k=\log_2\left(\det\left[\mathbf{I}+\mathbf{F}_k\mathbf{P}_k\text{diag}\left(\mathbf{a}_k\odot\mathbf{a}_k\right)\mathbf{P}_k^{H}\mathbf{F}^H_k\mathbf{R}_{\mathbf{z}_k\mathbf{z}_k}^{-1}\right]\right),\label{Mimo rate}
\end{equation}
where the covariance matrix of the effective noise is given by
\begin{equation}
\mathbf{R}_{\mathbf{z}_k\mathbf{z}_k}=\sum\limits_{\substack{i=1\\i\neq k}}^K\mathbf{F}_k\mathbf{P}_i\text{diag}\left(\mathbf{a}_i\odot\mathbf{a}_i\right)\mathbf{P}_i^{H}\mathbf{F}_k^{H}+\sigma_n^2\mathbf{I}.
\end{equation}
\vspace{-10mm}

\section{Rate Analysis}
In this section, we carry out the sum rate analysis of the proposed strategies combined with the RBD precoder. Let us consider the matrices $\mathbf{H}^{\left(k,j\right)}=\mathbf{H}_k^{\text{T}}\mathbf{P}^a_j$,$\mathbf{\Upsilon}^{\left(k,j\right)}=\underaccent{\ddot}{\mathbf{U}}_k^H \mathbf{H}^{\left(k,j\right)}\underaccent{\ddot}{\mathbf{V}}_j $ and $\mathbf{\tilde{\Upsilon}}^{\left(k,j\right)}=\underaccent{\ddot}{\mathbf{U}}_k^H \mathbf{\tilde{H}}^{\left(k,j\right)}\underaccent{\ddot}{\mathbf{V}}_j $ . Employing an RBD precoder leads us to the following received vector:
\vspace{-5mm}
\begin{align}
\mathbf{y}_k=&a_c s_c\mathbf{H}_k^{\text{T}}\mathbf{p}_c+\left(\underaccent{\ddot}{\mathbf{U}}_k\underaccent{\ddot}{\mathbf{\Psi}}_k+\mathbf{\tilde{\Upsilon}}^{\left(k,k\right)}\right)\text{diag}\left(\mathbf{a}_k\right)\mathbf{s}_k\nonumber\\
&+\sum\limits_{\substack{j=1\\ j\neq k}}^K\mathbf{\Upsilon}^{\left(k,j\right)}\text{diag}\left(\mathbf{a}_j\right)\mathbf{s}_j+\underaccent{\ddot}{\mathbf{U}}_k^H\mathbf{n}_k
\end{align}
For the Min-Max criterion, we have
\begin{equation}
R_{c,k,i}^{\left(RBD\right)}=\log_2\Bigg(1+\frac{a_c^2\lvert\mathbf{h}^{\left(k\right)^{\text{T}}}_{j}\mathbf{p}_c \rvert^2}{\rho^{\left(\text{RBD}\right)^2}_{i,k}+{\sum\limits_{\substack{j=1\\ j \neq k}}^K \sum\limits_{m\in\mathcal{M}_j}a_m^2\lvert \upsilon^{\left(k,j\right)}_{i,m-n_j}\rvert^2}+\sigma_n^2}\Bigg),\label{Rate Min Max stream}
\end{equation}
where $\rho_{i,k}^{\left(\text{RBD}\right)^2}=\sum\limits_{\substack{l \in \mathcal{M}_k}}a_l^2\lvert\psi_{l-n_k}^{\left(k\right)} u_{i,l-n_k}^{\left(k\right)}+\tilde{\upsilon}^{\left(k,k\right)}_{i,l-n_k}\rvert^2$. 
Then we set $R_{c,k}^{\left(\text{max}\right)}=\max_{i}R_{c,k,i}^{\left(RBD\right)}$ and use \eqref{MRC common rate} and \eqref{total ergodic rate} to obtain the performance of the Min-Max criterion. In a perfect CSIT scenario, we have $\mathbf{\Upsilon}^{\left(k,j\right)}=\mathbf{\hat{\Upsilon}}^{\left(k,j\right)}$ and $R_{c,k,i}^{\left(RBD\right)}$ is given by \eqref{Rate Min Max stream} with $\rho_{i,k}^{\left(\text{RBD}\right)^2}=\sum\limits_{\substack{l \in \mathcal{M}_k}}a_l^2\lvert\psi_{l-n_k}^{\left(k\right)} u_{i,l-n_k}^{\left(k\right)}+\hat{\upsilon}^{\left(k,k\right)}_{i,l-n_jk}\rvert^2$.

Let us now consider MRC for the $k$th user and evaluate the vector $\mathbf{r}_{k,j}$ with $j \in \mathcal{M}_q$ and the column index $t=j-n_q$. When $q=k$ the squared norm of vector $\mathbf{r}_{k,j}$ is reduced to:
\begin{equation}
\lVert\mathbf{r}_{k,j}\rVert^2=\lVert\underaccent{\ddot}\psi_t^{\left(k\right)}\underaccent{\ddot}{\mathbf{u}}^{\left(k\right)}_t+\underaccent{\ddot}{\mathbf{\tilde{H}}}_k^{\text{T}}\underaccent{\ddot}{\mathbf{v}}^{\left(k\right)}_{t}\rVert^2.\label{rki mrc}
\end{equation}

%


When $q\neq k$ the squared norm of $\mathbf{r}_{k,j}$ is given by
\begin{equation}
\lVert\mathbf{r}_{k,j}\rVert^2=E_{tr}\sum_{i=1}^{N_k}\left\lvert\sum_{l=1}^{N_t}\sum_{n=1}^{N_t}h_{i,l}^{\left(k\right)}\lambda_n^{\left(q\right)}\bar{v}_{l,n}^{\left(q\right)}\underaccent{\ddot}{v}_{n,t}^{\left(q\right)}\right\rvert^2,\label{MRC rkj RBD}
\end{equation}
where $\lambda_n^{\left(q\right)}=\left(\sqrt{E_{tr}\psi_n^{\left(q\right)}+N_r\sigma_n^2}\right)^{-1}$.
Substituting \eqref{rki mrc} and \eqref{MRC rkj RBD} in \eqref{SINR MRC} we get the SINR of the MRC criterion, which can be used in \eqref{MRC common rate} and \eqref{total ergodic rate} to obtain the achievable sum rate. Under perfect CSIT assumption, \eqref{rki mrc} is reduced to $\lVert\mathbf{r}_{k,j}\rVert^2=\lvert \underaccent{\ddot}{\psi}_t^{\left(k\right)}\rvert^2$.

Finally, we consider MMSEc and define $\mathbf{D}_k=\underaccent{\ddot}{\mathbf{U}}_k\underaccent{\ddot}{\mathbf{\Psi}}_k+\mathbf{\tilde{\Upsilon}}^{\left(k,k\right)}$, $\mathbf{J_k}=\text{diag}\left(\mathbf{a}_k\odot\mathbf{a}_k\right)$. In the case of MMSEc,  we have
\begin{equation}\small
\sum\limits_{i\in\mathcal{M}_k}a_i^2\lvert\mathbf{w}_k^{H}\mathbf{r}_{k,i}\rvert^2=\text{tr}\left(\mathbf{r}_{k,c}^H\mathbf{R}_{\mathbf{y}_k\mathbf{y_k}}^{-1}\mathbf{D}_k\mathbf{J}_k\mathbf{D}_k^H\mathbf{R}_{\mathbf{y}_k\mathbf{y_k}}^{-1}\mathbf{r}_{k,c} \right)\label{MMSE data}
\end{equation}
\begin{equation}\small
\sum\limits_{\substack{j=1\\j\notin \mathcal{M}_k}}^M a_j^2\lvert\mathbf{w}_k^{H}\mathbf{r}_{k,j}\rvert^2=\sum\limits_{\substack{j=1 \\j\neq k}}^K\text{tr}\left(\mathbf{r}_{k,c}^H\mathbf{R}_{\mathbf{y}_k\mathbf{y}_k}^{-1}\mathbf{\tilde{\Upsilon}}^{\left(k,j\right)}\mathbf{J}_k\mathbf{\tilde{\Upsilon}}^{\left(k,j\right)^H}\mathbf{R}_{\mathbf{y}_k\mathbf{y_k}}^{-1}\mathbf{r}_{k,c} \right)\label{MMSE interference}
\end{equation}
Substituting \eqref{MMSE data} and \eqref{MMSE interference} in \eqref{SINR general combiner} we obtain the SINR, which can be used in \eqref{MRC common rate} and \eqref{total ergodic rate} to obtain the sum rate.

\section{Simulations}

In this section we evaluate the performance of the proposed combining techniques in a RS-based MIMO system employing MMSE and RBD precoders. As reported in the literature \cite{Joham2005,StankovicHaardt2008}, these precoders outperform their ZF and BD counterparts by allowing small MUI to significantly reduce the power penalty associated with linear precoding. We set $\mathbf{G}_k=\mathbf{I}$ for the MMSE precoder, whereas the RBD precoder uses the receiver defined in \eqref{Precoder and receiver RBD} since we focus on evaluating the common combiners. We consider $N_t=12$
and $K=6$ for all simulations. Each user is equipped with 2 receive antennas. The inputs are
Gaussian distributed with zero mean and unit variance. Each coefficient of $\tilde{\mathbf{H}}$ follows a Gaussian distribution, i.e., $\sim\mathcal{CN}(0,\sigma^2_{e})$. We
consider additive white Gaussian noise and define $\text{SNR}\triangleq E_{tr}/ \sigma_n^2$ with $\sigma_n^2=1$ for all simulations. The ESR was
computed averaging 1000 independent channel realizations. For each
channel realization we obtained $\bar{R}_c$ and $\bar{R}_p$ employing 100 error
matrices. We use SVD over the channel
$\left(\mathbf{H}=\mathbf{U}\mathbf{\Psi}\mathbf{V}\right)$ and then set
$\mathbf{p}_c=\mathbf{v}_1$\footnote{Note that the optimization of the common precoder would further increase the sum rate performance. However, finding the optimum is a non convex problem and performing an exhaustive search would dramatically increase the computational complexity}. The power allocated to $s_c$ was found through exhaustive search in order to maximize the sum rate. Uniform power allocation is used across private users.

\begin{figure}[htb]
\begin{centering}
\includegraphics[scale=0.5]{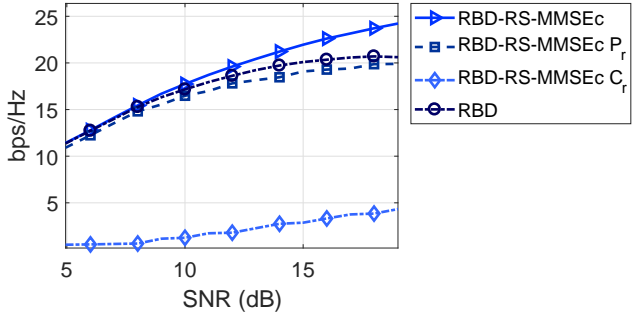}
\par\end{centering} \vspace{-0.75em}
\caption{Sum Rate performance under imperfect CSIT}\label{ImperfectCSIT}
\end{figure}

\begin{figure}[htb]
\begin{centering}
\includegraphics[scale=0.44]{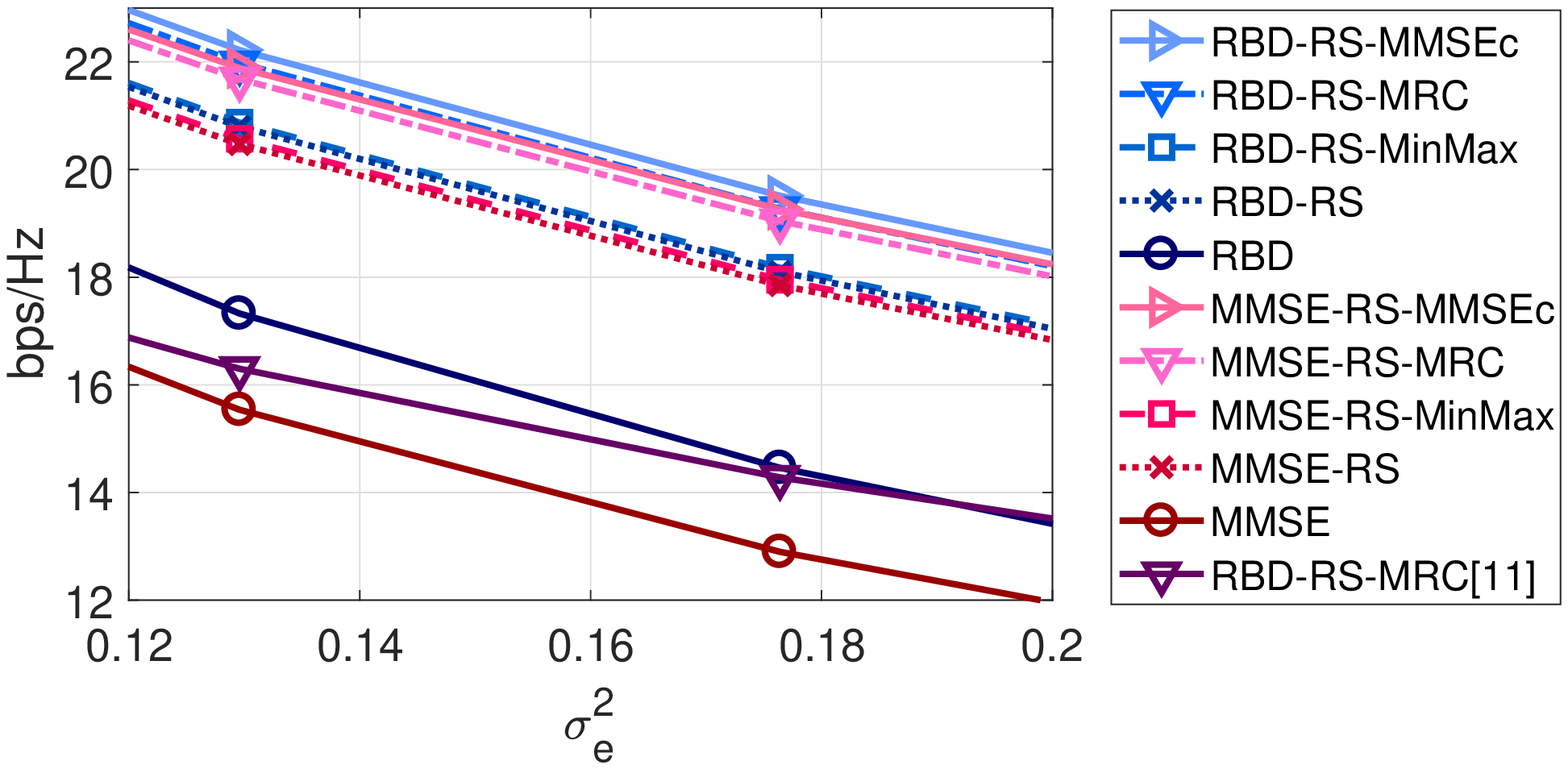}
\par\end{centering} \vspace{-0.75em}
\caption{Error variance VS Sum Rate performance}\label{VarErrZF}
\end{figure}

For the first simulation, we fixed the channel error variance to $\sigma_e^2=0.1$. Fig. \ref{ImperfectCSIT} shows the sum-private rate and the common rate of the RBD precoder with MMSEc, denoted by RBD-RS-MMSEc-Pr and RBD-RS-MMSEc-Cr respectively. The sum-private rate decreases up to 6\% when compared to the conventional RBD precoding since part of the transmit power is allocated to the common stream. However, the common rate attains up to 20\% of the conventional RBD rate, leading to an overall gain of the system performance. It is important to note that to obtain the gain an efficient power allocation scheme between common and private streams should be employed. The RS scheme deals partially with the MUI which is shown in Fig. \ref{ImperfectCSIT} where the common rate increases as the SNR grows.
Fig. \ref{VarErrZF} shows the performance of the proposed schemes as the estimation error increases. The conventional precoders are denoted by MMSE and RBD. MMSE-RS and RBD-RS denote the RS scheme without the common combiner. The best strategy from \cite{Flores2019} is represented by BD-RS-MRC. For this simulation we set the SNR to $20$ dB. The robustness of the system increases across all error variances when a common combiner is employed as shown in Fig. \ref{VarErrZF}. The figure shows that the proposed strategy outperforms the BD-RS-MRC scheme. The proposed MIMO RBD-RS-MMSEc attains a sum rate performance up to 34\% higher than conventional RBD. Moreover, MMSEc achieves the best performance among the combiners.

\begin{figure}[htb]
\begin{centering}
\includegraphics[scale=0.44]{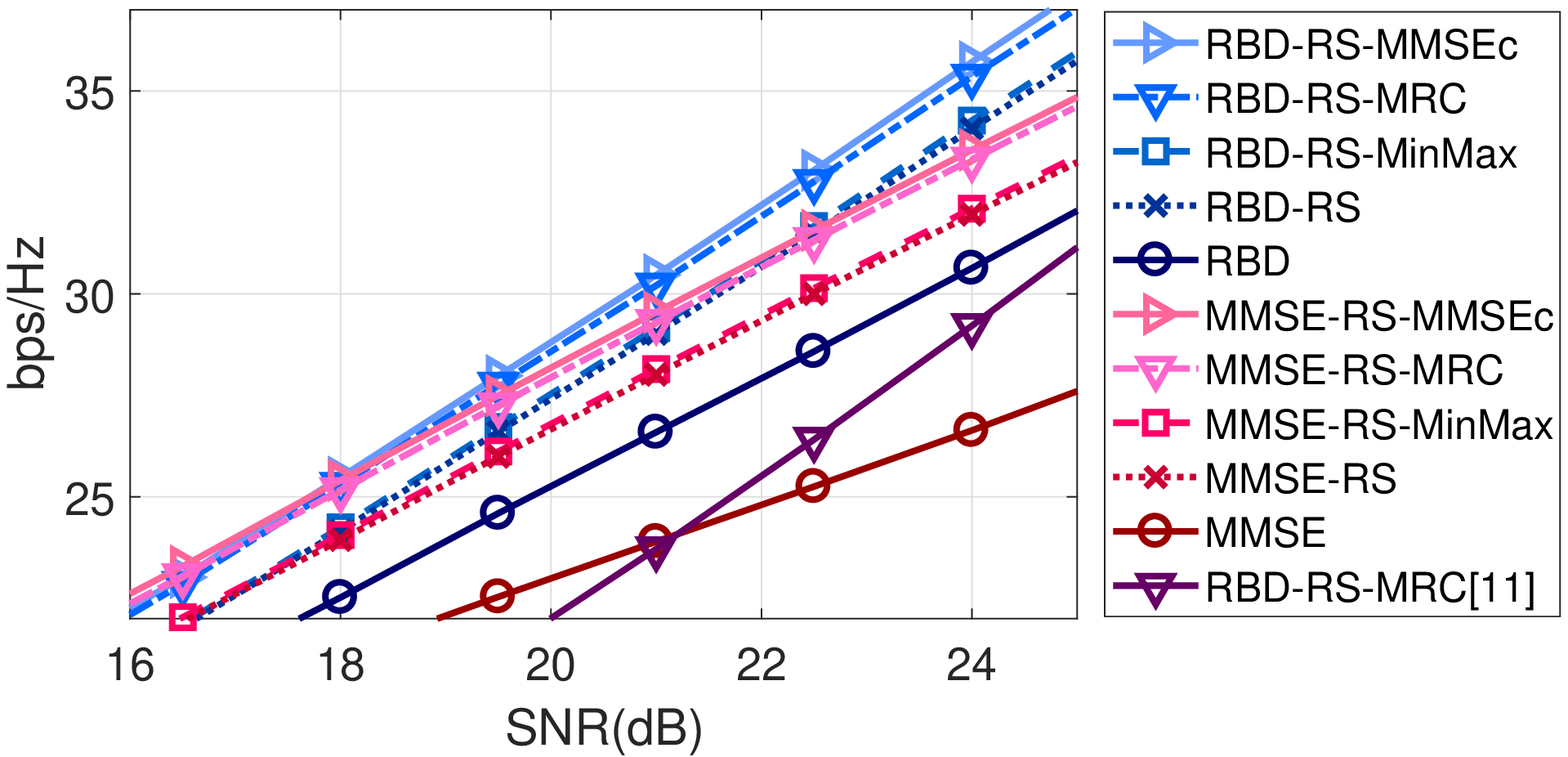}
\par\end{centering} \vspace{-0.75em}
\caption{Sum rate performance with imperfect CSIT.}\label{QualityZF}
\end{figure}

In the last example, we consider that the error in the channel estimate is reduced as the SNR increases, i.e. $\sigma^2_e=\xi
\left(E_{tr}/\sigma^2_n\right)^{-\alpha}$ with $\xi=0.94$ and $\alpha=0.6$. Fig.
\ref{QualityZF} shows that the use of combiners results in a higher sum rate than that of conventional schemes. The proposed MIMO RBD-RS-MMSEc obtains the best performance, which is up to 15\% when compared to conventional RBD precoding. Future work might consider massive MIMO systems \cite{mmimo,wence}


\section{Conclusion}
Simulation results show that employing a common stream significantly increase the overall performance of the system, contributing up to 20\% to the overall system sum rate. Furthermore, the proposed common stream combiners exploit the multipath propagation and the multiple antennas at the receiver to enhancing even more the performance of the common rate as shown by the simulations.  The RBD-RS-MMSEc shows an increase in the sum rate performance of more than 15\% when compared to conventional techniques. MMSEc also obtains the best performance among the combiners. Simulations have shown that the proposed stream combiners increase the robustness of the system under imperfect CSIT.


%





\ifCLASSOPTIONcaptionsoff
  \newpage
\fi

\bibliographystyle{IEEEtran}
\bibliography{CLref}
%

\end{document}